\documentclass[twocolumn, aps, prb]{revtex4}
\usepackage{bm}
\usepackage{graphicx}
\begin{document}
\title{Proximity-induced screening and its magnetic breakdown
in mesoscopic hybrid structures}

\author{Artem V. Galaktionov and Andrei D. Zaikin}
\affiliation{Forschungszentrum Karlsruhe, Institut f\"ur Nanotechnologie,
76021, Karlsruhe, Germany\\ I.E. Tamm Department of Theoretical Physics, P.N.
Lebedev Physics Institute, 119991 Moscow, Russia}
%\date{}

\begin{abstract}
We derive a general microscopic expression for the non-linear diamagnetic
current in a clean superconductor-insulator-normal metal structure
with an arbitrary interface transmission. In the absence of
electron-electron interactions in the normal metal the diamagnetic
response increases monotonously with decreasing temperature showing
no sign of paramagnetic reentrance down to $T=0$.
We also analyze the magnetic breakdown of proximity-induced Meissner
screening. We demonstrate that the magnetic breakdown field should be
strongly suppressed in the limit of small interface transmissions
while the linear diamagnetic
current does not depend on the transmission of the insulating barrier
at low enough temperatures.

\vspace{1.4cm}

\end{abstract}
\maketitle

\section{Introduction}
Mesoscopic hybrid structures composed of superconducting (S) and normal (N)
metals demonstrate a reach variety of intriguing physical phenomena. Many of
them have recently been extensively investigated -- both experimentally and
theoretically -- and received adequate interpretation within the framework of
the quasiclassical theory of superconductivity, see, e.g., Refs.
\onlinecite{lam,bel} for a recent review.

One of the remaining puzzles in the field concerns the paramagnetic reentrance
phenomenon observed in silver-niobium \cite{mot1,mot5} and gold-niobium
\cite{mot2} proximity cylinders in the limit of very low temperatures.
This reentrance behavior is in a clear contradiction with earlier theoretical
predictions \cite{zai} as well as with the results of more recent studies
\cite{bel2} demonstrating that diamagnetism in SN proximity cylinders should
progress {\it monotonously} with decreasing temperature. Several theoretical
explanations of this reentrance phenomenon have been put forward
\cite{BI,Zurich,Maki}. However, a detailed comparison with experiments carried
out in Refs. \onlinecite{mot5,mot2} indicates that none of these
explanations is able to correctly reproduce the absolute value or the
temperature dependence of the observed paramagnetic reentrance effect. One of
the explanations has also been subject to theoretical debate \cite{Zurich2}.

It is important to emphasize that, while the authors \cite{Zurich,Maki} invoke
additional assumptions about the form of electron-electron  interactions in
the N-metal, the work \cite{BI} operates with the standard model for SN
proximity systems which assumes no interaction between electrons in the normal
layer. This model is usually very well described by means of the standard
approach based on the quasiclassical Eilenberger equations. In order to test
the conjecture \cite{BI}, that paramagnetic reentrance could be missing within
the standard quasiclassical formalism because of an additional contribution
from the low energy ``glancing'' states, one needs to go beyond quasiclassics
and analyze the problem within a more general approach. Below we will use the
microscopic Gor'kov equations and evaluate the screening current by
constructing the full Green functions for the problem in question. We will
then compare our result with that derived from the Eilenberger equations. This
comparison confirms the applicability of the standard quasiclassical technique
and demonstrates that no paramagnetic reentrance can occur in SN proximity
cylinders in the absence of electron-electron interactions in the N-layer.

Another interesting property of the same structures is the so-called magnetic
breakdown of proximity-induced Meissner screening. This effect occurs at  a
certain value of the external magnetic field $H_{\rm b}$, above which the
diamagnetic expulsion of the field from the normal layer turns out to be
energetically unfavorable \cite{mot3,fau,bel2} although the proximity effect
itself is not yet suppressed. In the absence of impurities in the N-layer and
for the case of a perfectly transparent interface between S- and N-metals the
breakdown field $H_{\rm b}(T)$ was evaluated in Ref. \onlinecite{fau}. Since
at sufficiently low temperatures the {\it linear} current response of SN
proximity cylinders does not depend on the transmission of the SN interface
\cite{hig} one could assume that at $T \to 0$ the result \cite{fau} for
$H_{\rm b}$ should also apply at arbitrary transmissions. However, this turns
out not to be the case. As we will show below, the breakdown field scales
linearly with the interface transmission and, hence, $H_{\rm b}$ should be
strongly suppressed in the limit of small transmissions. This effect makes it
possible to extract information about the interface transmission from the
measurements of the magnetic breakdown field  $H_{\rm b}$.

Our paper is organized as follows. In section II  we will present a general
derivation of the non-linear screening current in superconductor-normal metal
proximity systems with arbitrary interface transmissions. This derivation is
carried out on the basis of the microscopic Gor'kov equations and does not
involve energy-integrated quasiclassical Eilenberger functions. The effect of
magnetic breakdown of proximity-induced screening is analyzed in section III.
Applicability of the quasiclassical approach for the problem in question is
discussed in section IV.

\section{Nonlinear diamagnetic response}
We will consider a clean SIN structure and assume it to be
uniform along the directions parallel to the interfaces
(coordinates $y$ and $z$). The normal metallic layer (N) and the bulk
superconductor (S) are located respectively at $0<x\leq d$ and $x<0$.
Our analysis will be based
on the Gor'kov equations of the microscopic theory of superconductivity
\cite{AGD}. After the Fourier transformation of the
normal ($G$) and anomalous ($F^+$) Green functions with respect to $y$ and $z$,
$$ G_{\omega_n}(\bm{r}, \bm{r'})=
\int\frac{d^2 \bm{k_\parallel}}{(2\pi)^2} G_{\omega_n}
(x,x',\bm{k_\parallel})e^{i \bm{k_\parallel}(\bm{r_\parallel}-
\bm{r'_\parallel})},
$$
the Gor'kov equations take the following standard form
\begin{equation}
{\small \left( \begin{array}{cc} i\omega_n -\hat H & \Delta(x)\\ \Delta^*(x)&
i\omega_n +\hat H_c\end{array}\right)\left( \begin{array}{c} G_{\omega_n}
(x,x',\bm{k_\parallel})\\F^+_{\omega_n} (x,x',\bm{k_\parallel})
\end{array}\right)= \left(\begin{array}{c} \delta(x-x')\\ 0\end{array}
\right)} \label{start}
\end{equation}
Here $\omega_n=(2n+1)\pi T$ is the Matsubara frequency, and $\Delta(x)$ is the
superconducting order parameter. Below we will choose $\Delta (x)=\Delta$ for
$x<0$ and $\Delta (x)=0$ otherwise. The Hamiltonian $\hat H$ in Eq.
(\ref{start}) is defined as
\begin{equation}
\hat H=-\frac{1}{2m}\frac{\partial^2}{\partial x^2}+ \frac{\bm{\tilde
k^2_\parallel}}{2m}-\epsilon_F +V(x), \label{H}
\end{equation}
where $\bm{\tilde k_\parallel}= \bm{k_\parallel}-
\frac{e}{c}\bm{A_\parallel}(x)$, $\epsilon_F$ is Fermi energy and
the term $V(x)$ accounts for the boundary potentials. The Hamiltonian
$\hat H_c$ is obtained from $\hat H$ (\ref{H}) by inverting the sign
of the electron charge $e$.

In what follows we shall neglect the square term $\propto \bm{A_\parallel}^2$
in Eq. (\ref{H}). This approximation applies in the range of magnetic fields
$H\ll \Phi_0 k_F/d$, where $k_F$ is the Fermi momentum and $\Phi_0=\pi c/e$ is
the superconducting flux quantum. We will assume that the London penetration
depth of the bulk superconductor is small and neglect the magnetic field
inside the superconductor. The latter assumption allows to set $\bm{A}(x\leq
0)=0$.

The next standard step is to decompose the Green functions into the product of
quickly oscillating terms $\exp(\pm ik_x x)$ and the two component envelope
functions  $\overline{\varphi}_\pm(x)$ changing at scales much longer
than the Fermi wavelength. One has
\begin{eqnarray}
\left( \begin{array}{cc} i\omega_n -\hat H & \Delta(x)\\ \Delta^*(x)&
i\omega_n +\hat H_c\end{array}\right)\overline{\varphi}_\pm(x)e^{\pm ik_x x}
\nonumber
\\
\simeq e^{\pm ik_x x}\left( \begin{array}{cc} i\omega_n -\hat H^{a}_\pm &
\Delta(x)\\ \Delta^*(x)& i\omega_n +\hat H^{a}_{\pm c}\end{array}\right)
\overline{\varphi}_\pm(x),
\end{eqnarray}
where we defined $k_x=\sqrt{k_F^2-k_\parallel^2}$ and
\begin{equation}
\hat H^{a}_\pm=\mp
iv_x\partial_x-\frac{e}{c}\bm{A_\parallel}(x)\bm{v_\parallel}.\label{si}
\end{equation}
A convenient choice of the gauge for the problem in question is
$\bm{A}=(0,A(x),0)$. Below we will make use of this gauge and
proceed along the lines with our previous analysis \cite{gz}.

Let us fix the coordinate $x'$ within the normal metal. In this case for $x<0$
the general solution of Eq. (\ref{start}) can be written in the form
\begin{eqnarray}
\left( \begin{array}{c} G_{\omega_n} (x,x')\\F^+_{\omega_n}
(x,x')\end{array}\right)=\left( \begin{array}{c} 1\\-i
\end{array}\right)e^{\Delta x/v_x}e^{ik_x x} s_1(x') \label{ls}  \\+
\left( \begin{array}{c} 1\\i
\end{array}\right)e^{\Delta x/v_x}e^{-ik_x x} s_2(x').\nonumber
\end{eqnarray}
Here we   restricted ourselves to small Matsubara frequencies
$\omega_n \ll \Delta$. This is sufficient in the most interesting physical
limit $T \ll\Delta$ and $d \gg \xi_0 \sim v_F/\Delta$ which we only consider
below.

The general solution of Eq. (\ref{start}) for $x>0$ reads
\begin{eqnarray}
\left( \begin{array}{c} G_{\omega_n} (x,x')\\F^+_{\omega_n}
(x,x')\end{array}\right)=  \left( \begin{array}{c} \tilde G_{\omega_n} (x,x')
\\0 \end{array}\right)+\label{rs} \\
+\left( \begin{array}{c} \varphi_+(x)e^{ik_x x}f_1(x') + \varphi_-(x)e^{-ik_x
x}f_2(x')
\\ \varphi_-(x)e^{ik_x x}f_3(x') +\varphi_+(x)e^{-ik_x x}f_4(x')
\end{array}\right) .\nonumber
\end{eqnarray}
Here the functions
\begin{eqnarray}
\varphi_+(x)=\exp\left(-\frac{\omega_n}{v_x}x+i\int_0^x
\frac{ek_y}{ck_x}A(x_1)dx_1 \right), \nonumber\\
\varphi_-(x)=\frac{1}{\varphi_+(x)}.\label{varph}
\end{eqnarray}
obey the equations $\hat H^{a}_\pm\varphi_\pm(x)=0$ and the first
term in the right hand side  of (\ref{rs}) with
\begin{equation}
\tilde G_{\omega_n}(x,x')=-\frac{i}{v_x}\left\{ \begin{array}{c}
\varphi_+(x)\varphi_-(x')e^{ik_x(x-x')} \quad\text{if} \:x>x'
\\  \varphi_-(x)\varphi_+(x')e^{ik_x(x'-x)} \quad\text{if} \:x<x'\end{array}
\right.
\end{equation}
represents the particular solution of Eq. (\ref{start}) for $x>0$. What
remains is to determine six functions $s_{1,2}(x')$ and $f_{1-4}(x')$ in Eqs.
(\ref{ls},\ref{rs}). This is done with the aid of the boundary conditions
imposed on both sides of the normal layer.

First we consider the SN interface. Matching the wave functions
on both sides of this interface, respectively $A_1 \exp(ik_{1x}x)+
B_1 \exp(-ik_{1x}x)$ and
$A_2 \exp(ik_{2x}x)+ B_2 \exp(-ik_{2x}x)$, is performed in a standard
manner (see, e.g., Ref. \onlinecite{LL}):
\begin{eqnarray}
A_2=\alpha A_1+\beta B_1,\: B_2=\beta^*A_1 +\alpha^* B_1,\nonumber
\\
 |\alpha|^2-|\beta|^2=1, \label{scatt}
\end{eqnarray}
where the reflection and transmission coefficients are equal to
\begin{equation}
R=\left|\frac{\beta}{\alpha}\right|^2,\quad D=1-R=\frac{1}{|\alpha|^2}.
\label{scatt2}
\end{equation}
Applying Eqs. (\ref{scatt}) to the two-component vectors $(G,F^+)$ we get four
linear equations for six unknown functions in Eqs. (\ref{ls},\ref{rs}). The
remaining two equations are derived at the interface between vacuum and the
normal metal ($x=d$) where we assume complete specular reflection of the
electrons. This boundary condition trivially yields
\begin{equation}
G(d,x')=F^+(d,x')=0.
\end{equation}
The above six linear equations uniquely define the Green function of our
system. Technically the calculation is similar to that presented in Ref.
\onlinecite{gz}, therefore we will not go into further details here. The
resulting expression for the Green function in the N-metal ($0<x\leq d$) takes
the form
\begin{eqnarray}
\label{G} && G_{\omega_n} (x,x,\bm{k_\parallel})=-\frac{1}{v_x}\left[
\frac{i\sinh \chi +\frac{2\sqrt{R}}{1+R}\sin(\gamma )}{\cosh \chi
+\frac{2\sqrt{R}}{1+R} \cos(\gamma ) }\right],\label{sno}\\ && \chi
=\frac{2\omega_n d}{v_x}-2i\int_0^d\frac{ek_y}{ck_x}A(x)d x,\quad \gamma =2k_x
d+\text{arg}\,\frac{\beta}{\alpha^*}.\nonumber
\end{eqnarray}
Of interest for us here is the current density in the
$y$-direction. It reads
\begin{equation}
j(x) =\frac{4e}{m}T\sum_{\omega_n>0}\int_{|k_\parallel|<k_F} \frac{d^2
\bm{k_\parallel}}{(2\pi)^2} k_y \text{Re}\,G_{\omega_n}
(x,x,\bm{k_\parallel}).\label{real}
\end{equation}
Note, that the terms $\sin\gamma$ and $\cos\gamma$ in Eq. (\ref{G}) are
quickly oscillating functions of $\bm{k_\parallel}$. Integration over the
momentum directions in (\ref{real}) is equivalent \cite{gz} to
averaging over the angle $\gamma$. Performing this averaging we
arrive at a spatially homogeneous screening current in the N-layer
\begin{equation}
j =\frac{4 e k_F^2 T}{\pi^2}\sum_{\omega_n>0}\int_0^{\pi/2}d\theta
\int_0^{\pi/2}d\varphi \sin^2\theta\cos\varphi\text{Im}\,G(\chi ).\label{f1}
\end{equation}
The function $G$ in Eq. (\ref{f1}) depends on the complex variable $\chi
=\chi_1 -i\chi_2$, where
\begin{equation}
\chi_1 =\frac{2\omega_n d}{v_F\cos\theta},\;\;\;\;
\chi_2=\phi\tan\theta\cos\varphi
\end{equation}
and
\begin{equation}
\phi=\frac{2\pi}{\Phi_0}\int_0^d A(x)d x.
\end{equation}
The function $G(\chi )$ is $\pi$-periodic in $\chi_2$.
Within one period it is defined
on a strip $-\pi/2<\chi_2 <\pi/2$ with the cut going from $(0,-\arcsin t)$
to $(0,\arcsin t)$. For such values of $\chi_2$ one has
\begin{eqnarray}
G=\frac{\sinh \chi }{\sqrt{t^2(\theta )+\sinh^2 \chi }},
\label{f2}
\end{eqnarray}
where $t(\theta)=D(\theta)/(1+R(\theta))$.
Eqs. (\ref{f1})-(\ref{f2}) fully determine the non-linear diamagnetic
response of a clean SIN system to an externally applied magnetic field $H$.

It is easy to check that our result reduces to the already known ones in the
corresponding limits. For instance, in the case of a perfectly transmitting SN
interface ($R=1-D=0$) Eqs. (\ref{f1})-(\ref{f2}) reproduce the non-linear
response derived in Ref. \onlinecite{zai}. Another important limit is that
of a small external field. In this case one can linearize the function $G$
(\ref{f2}) in $\phi$ and find
\begin{equation}
j=-\frac{e k_F^2 T \phi}{\pi}\sum_{\omega_n>0}\int_0^{\pi/2}
\frac{d\theta\sin^3\theta t^2(\theta)\cosh \chi_1}{
\cos\theta\left[t^2(\theta)+\sinh^2\chi_1 \right]^{3/2}},
\label{japan}
\end{equation}
This result was obtained in Ref. \onlinecite{hig} with the aid of a
different approach. It
is interesting that at low temperatures $T\ll tv_F/d$ the transmission and
reflection coefficients drop out and for any nonzero $t$ the result
(\ref{japan}) reduces to a simple formula
\begin{equation}
j=-\frac{e k_F^2 v_F \phi }{6\pi^2d},\label{jlt}
\end{equation}
which was initially derived \cite{zai} for $t=1$. At higher temperatures $t$ does not
anymore drop out of the final result. For instance, in the limit
$T\gg v_F/d$ we obtain
\begin{equation}
j=-\frac{e k_F^2 v_F t_0^2 \phi }{2\pi^3 d}\left(\frac{v_F}{Td}\right)e^{-4\pi
T d/v_F},
\end{equation}
where $t_0$ is the value of $t(\theta)$ at $\theta=0$. In the case of low
transmissions $t_0\ll 1$ there exists an additional temperature
interval $t_0v_F/d\ll
T \ll v_F/d$, where the current shows a power-law dependence on temperature:
\begin{equation}
j=-\frac{7\zeta(3)e k_F^2 v_F \phi }{64\pi^4 d } \left(\frac{v_F}{Td}\right)^
2 \int_0^{\pi/2}d\theta \cos^2\theta\sin^3\theta t^2(\theta). \label{pl}
\end{equation}

It is important to emphasize that the screening current in the
N-metal is {\it uniform} in space and, hence, is a nonlocal function of
the vector potential. Combining this result with the Maxwell
equations one easily arrives at the self-consistency condition for
the ``phase'' $\phi$
\begin{equation}
\frac{\phi c}{e}=Hd^2+\frac{8\pi}{3c}j(\phi )d^3. \label{self}
\end{equation}
In the linear response regime the current density $j$ can be expressed in the
form $j=-c^2\phi /8\pi e\lambda_N^2(T) d$.  As soon as the effective length
$\lambda_N(T)$ becomes small, $\lambda_N(T)\ll d$, the non-locality of the
screening current turns out to be important. For instance, at $T \to 0$ we
have $j=-3Hc/8\pi d$, the magnetic field penetrating into the normal layer
becomes over-screened \cite{zai}, $B(x)=3Hx/(2d)-H/2$, and the average
magnetization of the N-layer reads $4\pi {\cal M}=-3H/4$.

\section{Magnetic breakdown}

The above Meissner state of our hybrid structure is thermodynamically
favorable only in relatively small magnetic fields. At higher values of $H$
the diamagnetic screening current gets suppressed and the magnetic field can
freely penetrate into the N-layer. This magnetic breakdown effect was
experimentally studied by Mota and coworkers \cite{mot3} and was also
addressed theoretically \cite{bel2,fau} in the case of
perfectly transparent SN interfaces. In this section we will consider the
magnetic breakdown of the Meissner effect in clean proximity structures with
an arbitrary transmission of the SN interface. We will closely follow the
strategy adopted in Ref. \onlinecite{fau}.

The free energy $F$ of the current state (normalized per unit surface)
can be recovered by means of the standard procedure which amounts
to integrating the current (\ref{f1}) over an effective
``coupling constant'' $\lambda$
\begin{equation}
F(\phi )=-(\phi /2e)\int_{0}^{1}j(\lambda \phi)d \lambda .
\label{FE}
\end{equation}
In the presence of an external magnetic field we have to
minimize the Gibbs free energy
\begin{equation}
{\cal G}(T,H)=F(\phi )+\frac{1}{8\pi}\int_0^d dx(\partial_x A(x)-H)^2.
\label{Gibbs}
\end{equation}
In the limit of low external fields $H < H_{\rm min}$ the diamagnetic solution
described in the previous section is the only possible one. On the contrary,
at high magnetic fields $H >H_{\rm max}$ this solution cannot be realized. In
this case another solution of the Maxwell equations -- which describes full
penetration of the magnetic field into the N-layer -- takes over. In the
intermediate regime $H_{\rm min}<H <H_{\rm max}$ the diamagnetic and
non-magnetic solutions may coexist. The first order phase transition between
these two states occurs in this intermediate regime at a certain value of the
magnetic field $H=H_{\rm b}$ implying the magnetic breakdown of Meissner
screening. Since this breakdown field $H_{\rm b}$ can be sufficiently large,
the full nonlinear expression for the diamagnetic response, Eqs.
(\ref{f1})-(\ref{f2}), should be taken into account.

First let us briefly consider the case of a perfectly transparent SN interface
$t\equiv 1$ and re-derive the results \cite{fau}. The
``supercooling'' and ``superheating'' fields $H_{\rm min}$ and $H_{\rm max}$
can easily be estimated from Eq. (\ref{self}). Combining this equation with
the expression for the screening current $j=-c^2 \phi /8\pi e\lambda_N^2(T) d$
one readily finds $\phi \simeq 3\pi H\lambda_N^2(T)/\Phi_0$. Since the
diamagnetic solution is possible only for small $\phi \ll 2\pi$, one obtains
\begin{equation}
H_{\rm max} \sim \Phi_0/\lambda_N^2(T).
\label{max}
\end{equation}
For a non-magnetic solution the screening current is suppressed $j \approx 0$,
in which case from Eq. (\ref{self}) we find $\phi \simeq \pi Hd^2/\Phi_0$. The
vanishing screening current implies strong dephasing of Andreev states in the
N-layer. This dephasing effect occurs for $\phi \gg 2\pi$. Hence, we get
\begin{equation}
H_{\rm min} \sim \Phi_0/d^2.
\label{min}
\end{equation}

Now let us evaluate the breakdown field $H_{\rm b}$.
 The Gibbs energy of a non-magnetic state ${\cal G}_{nm}$ is minimized by the
equation $\partial_x A=H$. Making use of Eqs. (\ref{f1})-(\ref{f2}) and
(\ref{FE},\ref{Gibbs}) together with the inequality $\phi \gg 2\pi$ one finds
\begin{equation}
{\cal G}_{nm}=\frac{k_F^2 T}{\pi}\sum_{\omega_n>0}\int_0^1\ln\left[
1+\exp\left(-\frac{4\omega_n d}{\mu v_F}\right)\right]\mu d\mu ,\label{cg}
\end{equation}
where we denoted $\mu \equiv \cos\theta$.

The free energy of the diamagnetic state ${\cal G}_{dm}$ can also be
easily established. Assuming $\lambda_N(T)\ll d$ we obtain
\begin{equation}
{\cal G}_{dm}=3H^2d/32 \pi .
\label{dm}
\end{equation}

The breakdown field $H_{\rm b}$ is derived from the condition ${\cal
  G}_{dm}={\cal G}_{nm}$. In accordance with Ref. \onlinecite{fau} one gets
\begin{equation}
H_{\rm b} \simeq \frac{\sqrt{2}\Phi_0}{\pi\lambda_N(0)d}e^{-2\pi T d/v_F}
\label{hf}
\end{equation}
in the high temperature limit $T\gg v_F/d$ and
\begin{equation}
H_{\rm b} \simeq \frac{\Phi_0}{6\lambda_N(0)d}
\label{lt}
\end{equation}
in the opposite limit of low temperatures $T\ll v_F/d$. We observe that for
$\lambda_N(T)\ll d$ the breakdown field $H_{\rm b}$ is parametrically larger
(smaller) than $H_{\rm min}$ ($H_{\rm max}$), i.e. the phase transition indeed
occurs in the intermediate regime $H_{\rm min}<H <H_{\rm max}$. The condition
$\lambda_N(T^*)\sim d$ defines the temperature $T^*=(v_F/2\pi d)  \ln
(d/\lambda_N(0))$ below which the above picture remains valid. At higher
temperatures $T \gtrsim T^*$ a continuous and reversible cross\-over between
the two states is expected \cite{fau}.

Let us now turn to the case of a non-ideal SN interface $t
<1$. The estimate (\ref{min}) for the field
$H_{\rm min}$ remains the same while (\ref{max}) should
be modified. Below we will estimate $H_{\rm max}$ and obtain the expression for
$H_{\rm b}$ in three different temperature intervals.

{\it High temperatures.} In the limit $T\gg v_F/d$ Eqs. (\ref{f1})-(\ref{f2})
yield
\begin{equation}
j=-\frac{e k_F^2 v_F t_0^2 \phi }{2\pi^3 d}\left(\frac{v_F}{Td}\right)e^{-4\pi
T d/v_F}\exp\left[-\frac{v_F \phi^2}{2\pi T d } \right].
\end{equation}
Making use of this expression we evaluate the Gibbs energy and find $H_{\rm
b}$  which is again defined by Eq. (\ref{hf}) multiplied by the factor $t_0$.
Analogously, the estimate (\ref{max}) should now be multiplied by $t_0^2$ and
the whole picture remains consistent at $t_0 \gtrsim \lambda_N(T)/d$.

{\it Intermediate temperatures}.
In the limit $t_0\ll 1$ it is possible to realize the condition
$t_0v_F/d\ll T \ll v_F/d$. In this intermediate temperature range we obtain
\begin{eqnarray}
&&
j=-i\frac{ek_F^2v_F^2}{16\pi^4 T d^2}\int_0^{\pi/2}d\theta
\int_0^{\pi/2}d\varphi\cos^2\theta\sin^2\theta\cos\varphi \times\nonumber
\\&& t^2(\theta) \cos \chi_2 \left[\psi'\left(\frac{1+i\rho}{2}\right)-\psi'
 \left(\frac{1-i\rho}{2}\right )\right],
\label{full}
\end{eqnarray}
where $\psi(z)=\Gamma'(z)/\Gamma(z)$ is the digamma function and
\begin{equation}
\rho=
 \frac{v_F\cos\theta\sin\chi_2}{2\pi T d}.
\end{equation}
In the limit $\phi \ll T d/v_F$ we recover the result (\ref{pl}), while for $
\phi \gg T d/v_F$ the screening current {\it decreases} with
increasing $\phi$. For $T d/v_F\ll \phi \lesssim 1$ we get
\begin{equation}
j=-\frac{ek_F^2v_F}{4\pi^2 d \phi}\int_0^{1}\mu^2 t^2(\mu)d\mu .\label{ln}
\end{equation}

The estimate for $H_{\rm max}$ is obtained by combining Eqs. (\ref{pl}) and
(\ref{self}). In this regime $H_{\rm max}$ turns out to be smaller than
(\ref{max}) by the factor $\sim t_0^2v_F/Td$. The free energy of a
non-magnetic state is also established easily. Making use of the condition
$\phi \gg 2\pi$ and integrating the current (\ref{full}) over $\lambda$ in
(\ref{FE}) we obtain
\begin{equation}
F=\frac{k_F^2v_F}{8\pi^2 d}\int_0^1  d\mu\mu^2 t^2(\mu)\ln\kappa,
\label{nmt}
\end{equation}
where $\kappa=\gamma_0 \mu v_F/(2\pi Td)$ and $\ln\gamma_0 \simeq 0.577$ is
the Euler constant. The magnetic breakdown field $H_{\rm b}$ is obtained by
comparing Eqs. (\ref{dm}) and (\ref{nmt}). One finds
\begin{equation}
H_{\rm b}=\frac{\Phi_0}{\pi\lambda_N(0)d}\left[\int_0^1  d\mu\mu^2
t^2(\mu)\ln\kappa\right]^{1/2}. \label{brln}
\end{equation}
Thus, in the limit $t_0 \ll 1$
this field is strongly reduced as compared
to the case of perfectly transparent SN interfaces (\ref{lt}).

{\it Low temperatures.}
In the case $T\ll t_0v_F/d$ the diamagnetic current takes the form
\begin{equation}
j=-\frac{ek_F^2v_F}{\pi^3 d}\int_0^{\pi/2}d\theta \int_0^{\pi/2}d\varphi
\sin^2\theta\cos\theta\cos\varphi I(\chi_2 ,t).\label{l1}
\end{equation}
Here $I(\chi_2 ,t(\theta ))$ is a $\pi$-periodic function of $\chi_2
$. For $|\chi_2 |<\pi/2$ this function is defined as
\begin{equation}
I=\chi_2 -\text{sign}\,\chi_2 \,\Theta\,(|\sin \chi_2
|-t)\arctan\frac{\sqrt{\sin^2 \chi_2 -t^2}}{\cos \chi_2 },\label{nonl}
\end{equation}
where $\Theta(x)$ is the Heaviside step function. The function $I(\chi_2, t)$
is depicted in Fig. 1 for different values of $t$. We observe that
this function is linear in $\chi_2 \propto \phi$ only at $\phi\alt
t$. For
$t\ll \phi \lesssim 1$ we again recover the dependence (\ref{ln}).

The estimates for $H_{\rm max}$ and $H_{\rm b}$ follow from the above
expressions for the current. At $T \to 0$ the field $H_{\rm  max}$ turns out
to be by the factor $\sim t_0$ smaller than (\ref{max}) and $H_{\rm b}$ is
again defined by Eq. (\ref{brln}) with $\kappa \simeq 1/t_0$ in this case. The
above analysis remains consistent down to very small transmissions $t_0
\gtrsim \lambda_N(0)/d$.

The dependence of $H_{\rm b}$ on the transparency parameter $t_0$ in the limit
$T \to 0$ was also calculated numerically from Eqs. (\ref{l1},\ref{nonl}) and
the result is presented in Fig. 2. With a sufficient accuracy this dependence
can be approximated by a simple formula
\begin{equation}
H_{\rm b}(t_0)/H_{\rm b}(1)\simeq t_0.
\label{to}
\end{equation}

\begin{figure}
\includegraphics[width=3.3in]{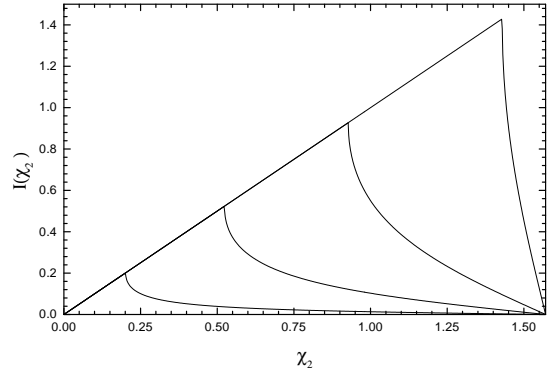}
\caption{The function $I(\chi_2)$ (\ref{nonl}) for different values of
$t=0.2, 0.5, 0.8, 0.99$ (left to right). The initial slope remains the
same for all curves.}
\end{figure}
\begin{figure}
\includegraphics[width=3.3in]{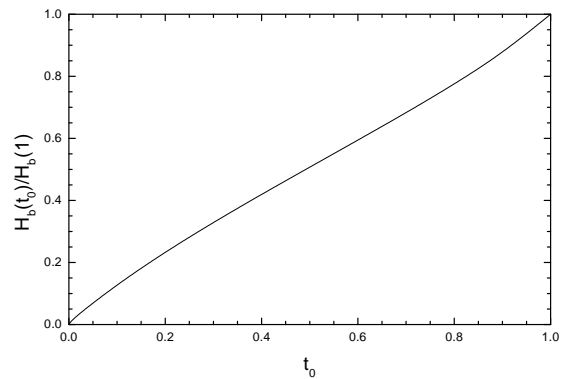}
\caption{The magnetic breakdown field $H_{\rm b}$ as a function of $t_0$
evaluated at $T=0$ for the angle-dependent transmission of the form
(\ref{thin}).}
\end{figure}

The above results for $H_{\rm b}(t_0)$ can be used to extract information
about the effective transmission of the SN interface from the measurements of
the breakdown field $H_{\rm b}$. A detailed comparison between
theoretical predictions (derived at $t=1$) and experimental data for
$H_{\rm b}$ was carried out in Refs. \onlinecite{fau,MMB}. It was found in
both papers that for comparatively clean samples the temperature
dependence of the
breakdown field is quite well described by a theory derived in the
clean limit. At the same time the amplitude of $H_{\rm b}$ turned out to be
somewhat smaller (typically by a factor $\sim 0.3\div 0.6$) than that
predicted in the ballistic case. One can attribute this discrepancy
to the effect of electron scattering on non-magnetic impurities
in the N-layer \cite{FN0}. It follows from the above analysis that
another possible reason for this suppression of $H_{\rm b}$
is a non-perfect transmission of the SN interface.

For instance, one can argue that the difference
between theory (\ref{lt}) and experiment by a factor $\approx 0.56$
found in Ref. \onlinecite{fau} at low $T$ could be due to
the effect of the interface transmission. Then from Fig. 2 one would recover
$t_0\approx 0.55$. In the most relevant case of a thin effective potential
barrier at the SN interface the angle-dependent transmission is defined as
\begin{equation}
t(\theta )=\frac{t_0\cos^2\theta}{1-t_0\sin^2\theta},
\label{thin}
\end{equation}
in which case the above estimate for $t_0$ will translate into the value of
the minimum interface reflectivity $R_{\rm min} \simeq 0.29$. Hence, one can
conclude that SN interfaces in the experiments \cite{mot3} were indeed highly
transmittive as it was already assumed before by a number of authors. Of
course, other factors, such as electron scattering on impurities in the
N-layer, differences in geometry and corrections due to the term $\bm{A}^2$
neglected in Eq. (\ref{si}), can cause additional minor discrepancies
\cite{FN} between theoretical and experimental values of $H_{\rm b}$ and can
slightly modify our estimate for $R_{\rm min}$. However, the above conclusion
can hardly be affected by these minor modifications. In fact,
the estimate for $R_{\rm min}$ could only increase if electron
scattering on impurities is taken into account in addition to the
effect discussed here.

\section{Applicability of quasiclassics and absence of paramagnetic reentrance}

Finally let us briefly address the applicability of the quasiclassical
Eilenberger approach for the problem in question. Solving the Eilenberger
equations in the S- and N-metals and matching these solutions at the SN
interface with the aid of the Zaitsev boundary conditions we have evaluated
the screening current in the N-layer of our proximity system. The
calculation is to much extent analogous to that presented in Appendix A of
Ref. \onlinecite{gz}, therefore we will not go into details here. Although
the result of this calculation does not exactly match with our general
expression (\ref{real}), it turns out to be {\it identical}
to one presented by Eqs. (\ref{f1})-(\ref{f2}). The latter result was obtained
from (\ref{real}) by means of averaging over all possible directions of the
Fermi momentum or, equivalently, over the angle $\gamma$. Similarly to
Ref. \onlinecite{gz} the difference
between the expressions for the current before and after this averaging is
small in the parameter $1/k_F d$. We also note that in the case of
highly transmitting interfaces, $R \to 0$, no averaging over $\gamma$
is needed and Eq. (\ref{real}) coincides exactly with the result of
the quasiclassical analysis \cite{zai}.

The above calculation was carried out in the limit of large spacial
dimensions $L_y$ and $L_z$ of our system respectively
in $y$ and $z$ directions. For finite
$L_{y,z}$ our results for the screening current acquire a small
correction. Its magnitude can easily be estimated by
rewriting the integral over the momentum directions in
Eq. (\ref{real}) as a sum over the conducting channels. Using
further the Euler-Maclaurin summation formula
\begin{equation}
\frac{1}{2}F(a)+\sum_{n=1}^\infty F(a+n)\approx \int_a^\infty F(x)
dx-\frac{1}{2}F'(a)\label{eul}
\end{equation}
we observe that the correction to our results due to finite $L_{y,z}$
is small as $1/\sqrt{\cal N}$,
where ${\cal N} \sim k_F^2 L_yL_z$ is the effective number of conducting
channels in the $x$-direction.

The parameters $1/k_F d$ and $1/\sqrt{\cal N}$, therefore,
control the accuracy of the quasiclassical Eilenberger approach in
our problem. Both these parameters are very small since they
contain the ratio between the Fermi
wavelength and at least one of the system dimensions. For instance,
for the systems studied in Refs. \onlinecite{mot1,mot5,mot2,mot3}
we estimate \cite{FN1}  $1/k_F d \lesssim 10^{-4}$ and
$1/\sqrt{\cal N} \lesssim 10^{-6}$. Hence, one can conclude that the
Eilenberger formalism remains very
accurate in this case and no significant physics is missing within this
formalism at any temperature and at any interface transmission.

The above estimates also indicate that within our model no
paramagnetic reentrance effect can
be expected in SN proximity cylinders. Indeed in
the limit $L_y \gg d$ the difference between the slab and cylinder geometries
is negligible and our results will be directly applicable to the latter
geometry if we identify $L_y$ with the circumference of the cylinder. Then the
correction from all electron states within the energy interval $\sim \Delta$
from the Fermi energy (obviously this correction also includes the
contribution of the low energy glancing states \cite{BI}) is small as $\sim
1/\sqrt{\cal N}$.

Although the conclusion about the absence of paramagnetic reentrance was
obtained here only in the clean limit and for specularly reflecting
interfaces, this conclusion should apply in the presence of impurities as
well. In fact, a purely ballistic situation appears to be most favorable for
the low energy glancing states. As soon as a finite electron mean free path
$l$ is introduced a proximity induced minigap $\sim v_F{\rm
min}(l^{-1},l/d^2)$ will develop in the normal metal \cite{Pilgram} and there
will be no glancing states in the system at all.

We also note that our present consideration does not include the contribution
of the electron states with energies smaller than $\epsilon_F-\Delta$. This
contribution to the current (which has nothing to do with superconductivity
and glancing states) can be roughly estimated with the aid of the standard
results for the persistent currents in normal metallic cylinders, see, e.g.,
Ref. [\onlinecite{Yuval}]. Making use of these results, in the limit of small
magnetic fields one can assume $I_{PC} \sim
e(v_F/L_y) k_F^{3/2} d^{1/2} L_z \phi$.
The relative magnitude of this contribution is also much smaller than the
proximity induced diamagnetic current studied here. At $T \to 0$ and for
relatively high interface transmissions one finds $I_{PC}/jdL_z \sim [d/k_F
L_y^2]^{1/2}$. For typical experimental parameters \cite{mot1,mot5,mot2} this
ratio is of order $\sim 10^{-4}$, i.e. in this case no noticeable correction
to our results could be expected either.

In summary, with the aid of the microscopic Gor'kov equations we have derived
a general expression for the non-linear diamagnetic current in a clean
superconductor-insulator-normal metal structure with an arbitrary interface
transmission. We have demonstrated that at low temperatures the
magnetic breakdown field $H_{\rm b}$ is suppressed linearly with the
transmission of the insulating barrier. This observation enables one
to obtain information about the quality of inter-metallic interfaces
from measurements of $H_{\rm b}$. We have also compared our results
with those derived within the quasiclassical Eilenberger approach and
found the latter approach to be extremely accurate for the systems
under consideration at all relevant temperatures down to $T=0$ and at
any interface transmission.

We would like to thank W. Belzig for useful remarks. This work
is part of the Kompetenznetz `` Functionelle Nanostructuren''
supported by the Landestiftung Baden-W\"urtemberg gGmbH.
One of us (A.V.G.) acknowledges support from the Alexander von Humboldt
Stiftung.


\begin{thebibliography}{}
\bibitem{FN0} It was demonstrated in Ref. \onlinecite{MMB} that a
quantitative theoretical description of the experimentally observed
diamagnetic susceptibility is possible only provided the effect
of a finite electron mean free path in the N-metal is taken into account.
\bibitem{FN} Using Eq. (\ref{hf}) multiplied by $t_0\approx 0.56$ one
can obtain an estimate for $H_{\rm b}$ at high temperatures
$T \gg v_F/d$. The corresponding result will coincide with
one presented by the solid curve in Fig.3
of Ref. \onlinecite{fau}. However, the experimental data in this high temperature
regime turn out to be smaller than this estimate by a
constant factor $\approx 0.65$.
\bibitem{FN1} In our estimates we use typical values
$d \sim 5 \mu$m, $L_y \sim 100 \mu$m and $L_z \sim 1$mm.
\end{thebibliography}
\end{document}